\def\bold#1{\setbox0=\hbox{$#1$}
     \kern-.025em\copy0\kern-\wd0
     \kern.05em\copy0\kern-\wd0
     \kern-.025em\raise.0433em\box0 }
\def\slash#1{\setbox0=\hbox{$#1$}#1\hskip-\wd0\dimen0=5pt\advance
       \dimen0 by-\ht0\advance\dimen0 by\dp0\lower0.5\dimen0\hbox
         to\wd0{\hss\sl/\/\hss}}
\documentstyle[12pt]{article}

\newlength{\dinwidth}
\newlength{\dinmargin}
\newcommand{\resection}[1]{\setcounter{equation}{0}\section{#1}}

\begin{document}
\def\lq{\left [}
\def\rq{\right ]}
\def\LL{{\cal L}}
\def\VV{{\cal V}}
\def\AA{{\cal A}}
\newcommand{\be}{\begin{equation}}
\newcommand{\ee}{\end{equation}}
\newcommand{\bea}{\begin{eqnarray}}
\newcommand{\eea}{\end{eqnarray}}
\newcommand{\nn}{\nonumber}
\newcommand{\dd}{\displaystyle}
\thispagestyle{empty}
\vspace*{4cm}
\begin{center}
  \begin{Large}
  \begin{bf} EFFECTIVE LAGRANGIAN FOR HEAVY AND LIGHT MESONS: RADIATIVE B
DECAYS.$^*$\\
  \end{bf}
  \end{Large}
  \vspace{1cm}
  \begin{large}
R. Casalbuoni\\
  \end{large}
Dipartimento di Fisica, Univ. di Firenze\\
I.N.F.N., Sezione di Firenze\\
  \vspace{5mm}
  \begin{large}
A. Deandrea, N. Di Bartolomeo and R. Gatto\\
  \end{large}
D\'epartement de Physique Th\'eorique, Univ. de Gen\`eve\\
  \vspace{5mm}
  \begin{large}
G. Nardulli\\
  \end{large}
Dipartimento di Fisica, Univ.
di Bari\\
I.N.F.N., Sezione di Bari\\
  \vspace{5mm}
\end{center}
  \vspace{2cm}
\begin{center}
UGVA-DPT 1993/04-816\\
BARI-TH/93-140\\
hep-ph/9304302\\
April 1993\\
\end{center}
\vspace{1cm}
\noindent
$^*$ Partially supported by the Swiss National Foundation
\newpage
\thispagestyle{empty}
\begin{quotation}
\vspace*{5cm}
\begin{center}
  \begin{Large}
  \begin{bf}
  ABSTRACT
  \end{bf}
  \end{Large}
\end{center}
  \vspace{5mm}
\noindent
We make use of the information obtained from semileptonic decays of $B$ and $D$
mesons, within an effective lagrangian description based on heavy quark theory
and on chiral expansion, to study the radiative decays $B \to K^\star
\gamma$ and $B_S \to \phi\gamma$, and the pair conversion processes
$B \to K e^+e^-$ and $B \to K^\star e^+e^-$. We discuss with care the required
extrapolations from zero recoils. We obtain a BR($B\to K^\star\gamma$) between
$1.3\times 10^{-5}$ and $4.1\times 10^{-5}$ and BR($B_s \to\phi\gamma$)
between $1.4\times 10^{-5}$ and $4.5\times 10^{-5}$, with errors mainly
arising from the present uncertainty in the Kobayashi-Maskawa element
$|V_{ts}|$. For the pair conversion processes we study in particular the lepton
effective mass distributions whose measurements would allow for an
understanding both long distance and short distance contributions to such
processes.
\end{quotation}
\newpage
\setcounter{page}{1}
\resection{Introduction}
Radiative $B$ decays into strange final states have received much
theoretical attention in the last few years both for their relevance in
the framework of the Standard Model and for their possible role in the
discovery of new physics. As to the former aspect, it should be observed
that the short distance effective hamiltonian describing the decay
$b\to s\gamma$ contains, through loop effects, information on the top
quark mass as well as on some poorly known elements of the
Cabibbo-Kobayashi-Maskawa matrix. The top quark appears as a virtual state
in a penguin-like diagram contributing to $b\to s\gamma$ and is
responsible of a large QCD enhancement in the branching ratio
\cite{bertolini}, a phenomenon previously noted in $s\to d\gamma$ decay
\cite{vasanti}, \cite{inami}. Also the latter aspect of interest, i.e.
the relevance of radiative $B$ decays for the discovery of new
effects beyond the Standard Model, has been deeply studied and in
particular we mention here some analyses in the framework of
supersymmetry \cite{bertolini2}
\par
In this letter we wish to analyze the following exclusive
processes
\bea
B &\to & K^\star\gamma \nn\\
B_s & \to & \phi \gamma
\eea
\be
B\to K e^+ e^-
\ee
\be
B\to K^\star e^+ e^-
\ee
that are described at the quark level by the above mentioned short
distance $b\to s\gamma$ hamiltonian plus long distance contributions
arising from $J/\psi-\gamma$ and $\psi^\prime-\gamma$ conversion
(for virtual photons).
\par
As shown in \cite{isgur}, in the limit of infinitely heavy $b$ quark
one can relate the short distance hadronic matrix elements for the
transitions (1.1)-(1.3) to the matrix elements of the weak currents between
a heavy and a light meson, by using the flavour and spin symmetries of the
Heavy Quark Effective Theory [HQET] \cite{gen}. It should be observed that
while the currents appearing in the hadronic matrix elements of (1.1) are
computed at $q^2=0$ , the predictions of HQET are
in general reliable for $q^2\approx q^2_{max}$, i.e. at zero recoil
momentum. In \cite {burdman} it has been shown
that such an extrapolation does not
generate large corrections and that there exists a special kinematical point
in the semileptonic decay $B\to\rho e\nu$ where one can make predictions
that are largely free from hadronic uncertainties; this idea has been
subsequently developed in \cite{donnell}.
\par
Our approach to the processes (1.1)-(1.3) is based on an effective chiral
theory including mesons containing one heavy quark that has been
recently developed in \cite{wise}, \cite{noi1}.
By this approach we are
able to compute processes (1.1)-(1.3) by an effective lagrangian
possessing flavour and spin symmetry in the heavy degrees of freedom,
and chiral symmetry in the light ones. Due to these symmetries we can
evaluate the amplitudes for the processes (1.1)-(1.3) in terms of
some constants appearing in the semileptonic decays of $B$ and $D$'s.
Using the results obtained in \cite{noi1} for these decays, we can
therefore make predictions for the rare neutral flavour changing
processes (1.1)-(1.3).
\par
We conclude this introduction by briefly reviewing some notations of the
effective chiral
field theory for heavy mesons \cite{wise}, \cite{noi1}. The $J^P=0^-$
and $1^-$ heavy $Q{\bar q}_a$ mesons are represented by a $4\times 4 $
Dirac matrix
\bea
H_a &=& \frac{(1+\slash v)}{2}[P_{a\mu}^*\gamma^\mu-P_a\gamma_5]\\
{\bar H}_a &=& \gamma_0 H_a^\dagger\gamma_0
\eea
where $a=1,2,3$ (for $u$, $d$, $s$), $v$ is the heavy meson velocity,
(for $u,d$ and $s$ respectively),
$P^{*\mu}_a$ and $P_a$ are annihilation operators satisfying
\bea
\langle 0|P_a| H_a (0^-)\rangle & =&\sqrt{M_H}\\
\langle 0|P^{*\mu}_a| H_a (1^-)\rangle & = & \epsilon^{\mu}\sqrt{M_H}
\eea
where $v^\mu P^*_{a\mu}=0$ and $M_H$ is the heavy meson mass.
The light pseudoscalar mesons are described by
\be
\xi=\exp{\frac{iM}{f_{\pi}}}
\ee
where
\be
{M}=
\left (\begin{array}{ccc}
\sqrt{\frac{1}{2}}\pi^0+\sqrt{\frac{1}{6}}\eta & \pi^+ & K^+\nn\\
\pi^- & -\sqrt{\frac{1}{2}}\pi^0+\sqrt{\frac{1}{6}}\eta & K^0\\
K^- & {\bar K}^0 &-\sqrt{\frac{2}{3}}\eta
\end{array}\right )
\ee
with $f_{\pi}=132 MeV$. Under the chiral group $SU(3)\otimes SU(3)$
the fields transform as
follows
\bea
\xi & \to  & g_L\xi U^\dagger=U\xi g_R^\dagger\\
\Sigma =\xi^2 & \to  & g_L\Sigma {g_R}^\dagger\\
H & \to  & H U^\dagger\\
{\bar H} & \to & U {\bar H}
\eea
where  $g_L$, $g_R$ are global $SU(3)$
transformations and $U$ depends on the space point $x$, the fields, $g_L$
and $g_R$.
\par
The vector meson resonances belonging to the low lying $SU(3)$ octet can
be introduced by using the hidden gauge symmetry approach,
where the $1^-$ particles are the
gauge bosons of a gauge local subgroup $SU(3)_{loc}$. Starting from chiral
$U(3)\otimes U(3)$, one gets a nonet of vector fields
$\rho_\mu$ that describe the particles $\rho$, $\omega$, $K^\star$ and
$\phi$ (with ideal mixing, i.e. $\phi={\bar s}s$).
In this way the HQET chiral lagrangian describing the
fields $H$, $\xi$, $\rho_\mu$ as well as their interactions, is given, at
the lowest order in light meson derivatives, as follows \cite{noi1}
\be
\LL=\LL_{0}+\LL_{2}
\ee
\be
\LL_{0}=\frac{f_{\pi}^2}{8}<\partial^\mu\Sigma\partial_\mu
\Sigma^\dagger > +i < H_b v^\mu D_{\mu ba} {\bar H}_a >
+i g <H_b \gamma_\mu \gamma_5 \AA^\mu_{ba} {\bar H}_a>
\ee
\bea
\LL_2&=& -\frac{f^2_{\pi}}{2}a <(\VV_\mu-
\rho_\mu)^2>+\frac{1}{2g_V^2}<F_{\mu\nu}(\rho)F^{\mu\nu}(\rho)> \nn\\
&+&i\beta <H_bv^\mu\left(\VV_\mu-\rho_\mu\right)_{ba}{\bar H}_a>\nn\\
&+&\frac{\beta^2}{2f^2_{\pi} a}<{\bar H}_b H_a{\bar H}_a H_b>+
i \lambda <H_b \sigma^{\mu\nu} F_{\mu\nu}(\rho)_{ba} {\bar H}_a>
\eea
where $\langle\cdots\rangle$ means a trace,
\bea
D_{\mu ba}&=&\delta_{ba}\partial_\mu+\VV_{\mu ba}
=\delta_{ba}\partial_\mu+\frac{1}{2}\left(\xi^\dagger\partial_\mu \xi
+\xi\partial_\mu \xi^\dagger\right)_{ba}\\
\AA_{\mu ba}&=&\frac{1}{2}\left(\xi^\dagger\partial_\mu \xi-\xi
\partial_\mu \xi^\dagger\right)_{ba}
\eea
\be
F_{\mu\nu}(\rho)=\partial_\mu\rho_\nu-\partial_\nu\rho_\mu+
[\rho_\mu,\rho_\nu]
\ee
Moreover
\be
\rho_\mu=i\frac{g_V}{\sqrt{2}}\hat\rho_\mu
\ee
where
\be
{\hat \rho}=
\left (\begin{array}{ccc}
\sqrt{\frac{1}{2}}\rho^0+\sqrt{\frac{1}{2}}\omega & \rho^+ & K^{\star +}\nn\\
\rho^- & -\sqrt{\frac{1}{2}}\rho^0+\sqrt{\frac{1}{2}}\omega & K^{\star 0}\\
K^{\star -}& {\bar K}^{\star 0} &\phi
\end{array}\right )
\ee
and
\be
a=2 \ \ \ \ \ \ \ \ \ \ \ \ \ \ \ g_V \approx 5.8
\ee
 from the KSRF relations.
One can also introduce explicit symmetry breaking terms as
illustrated in \cite{wise}, \cite{noi1}.
\par
We shall also introduce
positive parity $0^+$ and $1^+$ heavy mesons. As described in \cite{noi1}
we are interested in the states having the total angular momentum of the light
degrees of freedom equal to 1/2: $s_\ell=1/2$. They are described by a
$4\times 4$ Dirac matrix analogous to (1.4)
\be
S_a=\frac{1+\slash v}{2} \left[D_1^\mu\gamma_\mu\gamma_5-D_0\right]
\ee
where $D_1^\mu$ and $D_0$ annihilate a $1^+$ and $0^+$ $Q{\bar q}_a$
states respectively with a normalization analogous to (1.6), (1.7).
The complete lagrangian can be found in \cite{noi1}.
\resection{Hadronic matrix elements}
As shown in \cite{bertolini}, the decay $B\to K^\star\gamma$ is mediated at
short distances by electromagnetic penguin diagrams that generate an effective
hamiltonian given by
\be
H_\gamma=Cm_b\bar s\sigma^{\mu\nu}(1+\gamma_5)b F_{\mu\nu}~~~+~~~{\rm h.c.}
\ee
where we have neglected terms of order $m_s/m_b$. $F_{\mu\nu}$ is the
electromagnetic tensor, $\sigma_{\mu\nu}=\frac{i}{2}[\gamma_\mu,\gamma_\nu]$
and $C$ is given by
\be
C=\frac{G_F}{\sqrt{2}}\frac{e}{16\pi^2}V_{tb}V_{ts}^* F_2\left(\frac{m_t^2}
{m_W^2}\right)
\ee
where
\be
F_2(x)=r^{-16/23}\left\{{\bar F}_2(x)+\frac{116}{27}\left[\frac{1}{5}
\left(r^{10/23}-1\right)+\frac{1}{14}\left(r^{28/23}-1\right)\right]
\right\}
\ee
with $r=\alpha_s(m_b)/\alpha_s(m_W)$ and ${\bar F}_2(x)$ given by
\be
{\bar F}_2(x)=\frac{x}{(x-1)^3}\left[\frac{2x^2}{3}+\frac{5x}{12}-\frac{7}{12}
-\frac{3x^2-2x}{2(x-1)}\log x\right]
\ee
$F_2(x)$ is a smooth function of the top quark mass $m_t$ with values between
0.55 (at $m_t=90~GeV$) and 0.68 (for $m_t=210~GeV$).
\par
The hamiltonian (2.1) describes also the processes (1.2) and (1.3), where
a virtual photon is emitted; however in these cases long distance
contributions play a major role \cite{riaz}, as we shall discuss in
section 4. On the other hand for $B\to K^\star\gamma$
or $B_s \to \phi\gamma$ long distance effects can be
safely neglected \cite{deshpande}, \cite{colangelo}.
\par
The short distance hadronic matrix element relevant to the transition
$\bar B\to K^\star\gamma$ ($\bar B= B^-$ or ${\bar B}^0$) can be expressed as
follows:
\bea
\langle K^\star(p^\prime,\epsilon)|\bar s\sigma^{\mu\nu}(1+\gamma_5)b|{\bar B}
(p)\rangle  &=& i\Big\{A(q^2)\left[p^\mu\epsilon^{*\nu}-p^\nu\epsilon^{*\mu}-
i\epsilon^{\mu\nu\lambda\sigma}p_\lambda\epsilon^*_\sigma\right] \nn\\
&+& B(q^2)\left[p^{\prime\mu}\epsilon^{*\nu}-p^{\prime\nu}\epsilon^{*\mu}-
i\epsilon^{\mu\nu\lambda\sigma}p^\prime_\lambda\epsilon^*_\sigma\right] \nn\\
&+& H(q^2)(\epsilon^*\cdot p)\left[p^\mu p^{\prime\nu}-p^\nu p^{\prime\mu}-
i\epsilon^{\mu\nu\lambda\sigma}p_\lambda p_\sigma^\prime\right]\Big\}
\eea
\par
On the other hand, for the transition $B\to K\gamma$, that can occur only
with virtual photons, we have  the short distance hadronic matrix element
\be
\langle K(p^\prime)|\bar s\sigma^{\mu\nu}(1+\gamma_5)b|{\bar B}(p)\rangle
=iS(q^2)\left[p^\mu p^{\prime\nu}-p^\nu p^{\prime\mu}-
i\epsilon^{\mu\nu\lambda\sigma}p_\lambda p_\sigma^\prime\right]
\ee
Both in (2.5) and in (2.6) we have used the property
$\frac{i}{2}\epsilon^{\mu\nu\lambda\sigma}\sigma_{\lambda\sigma}\gamma_5=
-\sigma^{\mu\nu}$ that allows to express matrix elements of $\bar s
\sigma^{\mu\nu}b$ in terms of those of $\bar s\sigma^{\mu\nu}\gamma_5 b$ .
\par
At the lowest order in the derivatives of the pseudoscalar fields, the weak
tensor current between light pseudoscalar and negative parity heavy mesons
is as follows
\be
L_{\mu\nu}^a=i\frac{\alpha}{2}\langle\sigma_{\mu\nu}(1+\gamma_5)H_b
\xi^\dagger_{ba}\rangle
\ee
that has the same transformation properties of the quark current
${\bar q}^a\sigma^{\mu\nu}(1+\gamma_5)Q$. Together with (2.7) we also consider
the weak effective current (\cite{wise}, \cite{noi1}) corresponding to the
quark $V-A$ current ${\bar q}^a\gamma^\mu(1-\gamma_5)Q$, i.e.
\be
L_\mu^a=i\frac{\alpha}{2}\langle\gamma_\mu(1-\gamma_5)H_b\xi^\dagger_{ba}
\rangle
\ee
We put the same coefficient $i\alpha/2$ in (2.7) and (2.8) because, as a
consequence of the equations of motion of the heavy quark
$\left( \frac{1+\slash{v}}{2}b=b\right)$, we have, in the $b$ rest frame
\cite{isgur},
\be
\gamma^0 b=b
\ee
Therefore
\be
{\bar q}^a\sigma_{0i}(1+\gamma_5)Q=-i{\bar q}^a\gamma_i(1-\gamma_5)Q
\ee
and the effective currents $L_{\mu\nu}^a$ and $L_\mu^a$ have to satisfy,
in the heavy meson rest frame, the relation
\be
L_{0i}=-iL_i
\ee
\par
We also introduce the weak tensor current containing the light vector meson
$\rho^\alpha$ and reproducing ${\bar q}^a\sigma^{\mu\nu}(1+\gamma_5)Q$
\be
L_{1a}^{\mu\nu}=i\alpha_1\left\{g^{\mu\alpha}g^{\nu\beta}-\frac{i}{2}
\epsilon^{\mu\nu\alpha\beta}\right\}\langle\gamma_5H_b\left[
\gamma_\alpha(\rho_\beta-\VV_\beta)_{bc}-\gamma_\beta(\rho_\alpha-
\VV_\alpha)_{bc}\right]\xi^\dagger_{ca}\rangle
\ee
$L_{1a}^{\mu\nu}$ is related to the vector current $L_{1a}^\mu$ introduced
in \cite{noi1} to represent ${\bar q}^a\gamma^\mu(1-\gamma_5)
Q$ between light vector particles and heavy mesons:
\be
L_{1a}^\mu=\alpha_1\langle\gamma_5 H_b(\rho^\mu-\VV^\mu)_{bc}\xi^\dagger_{ca}
\rangle
\ee
In order to construct the tensor current we have imposed, similarly to (2.11)
the relation $L_{1a}^{0i}=-iL_{1a}^i$.
\par
Let us briefly discuss the coupling constants appearing in the previous
equations. $\alpha$ and $\hat\alpha$ are related to the leptonic
constants defined by
\be
\langle 0|{\bar q}^a\gamma^\mu\gamma_5Q|P_b(p)\rangle=i\frac{\alpha}
{\sqrt{m_P}}\delta_{ab}
\ee
\be
\langle 0|{\bar q}^a\gamma^\mu Q|S_b(p)\rangle=i\frac{\hat\alpha}
{\sqrt{m_S}}\delta_{ab}
\ee
where $P_b$ and $S_b$ are the $Q{\bar q}_b$ mesons with $J^P=0^-$ and
$0^+$ respectively. Both $\alpha$ and $\hat\alpha$ have a smooth
logarithmic dependence on $m_Q$ that we omit since it leads to
negligible numerical effects. From $QCD$ sum rules analysis, as
applied to $f_B$, we obtain \cite{colangelo2} $f_B=\alpha/\sqrt{m_B}\simeq
200~MeV$, which implies
\be
\alpha\simeq 0.46~~GeV^{3/2}
\ee
We assume here that $m_B$ is large enough so that $1/m_B$ is
negligible and one could deduce $\alpha$ by $\alpha\approx f_Bm_B^{1/2}$;
QCD sum rules have also been applied to the determination of $\alpha$ in
the $m_Q\to\infty$ limit; however in this case, large ($\approx 100\%$)
${\cal O}(\alpha_s)$ corrections \cite{broadhurst} make the results unreliable.
As to $\hat\alpha$, we take the results of an analysis
\cite{colangelo3} based on QCD sum rules:
\be
\hat\alpha\simeq \alpha\approx 0.46~GeV^{3/2}
\ee
We note that this value of $\hat\alpha$, obtained in the limit $m_Q\to\infty$
coincides, modulo logarithmic corrections, with the result obtained by
extracting $\hat\alpha$ from $f_{B(0^+)}$ i.e.
$\hat\alpha\approx f_{B(0^+)}\sqrt{m_{B(0^+)}}$.
\par
Finally the constant $\alpha_1$ in (2.13) appears in
combination with other coupling constants and will be discussed in the next
Section.
\resection{The decays $B\to K^\star\gamma$ and $B_s\to \phi\gamma$}
To compute $B\to K^\star\gamma$ we consider a polar diagram (with a heavy meson
intermediate $1^+$ and $1^-$ state between the current and the $B K^\star$
system) and a direct term.
We assume that the effective lagrangian and the effective tensor
currents of the previous Section provide a reliable way to describe
the process only for large values of $q^2$, i.e. $q^2\approx q^2_{max}=(m_B-
m_{K^\star})^2$, since in writing strong and weak effective couplings we
have neglected higher order derivatives of the light fields. Moreover near the
zero recoil point the residual heavy meson momentum is small, which is a basic
condition of HQET. Following
\cite{noi1} we assume polar dependence in $q^2$ (with pole mass suggested
by dispersion relations), which represents quite well the data
in semileptonic heavy meson decays (\cite{stone}). By using the Feynman
rules for the heavy meson  chiral lagrangian given in refs.\cite{wise},
\cite{noi1} we obtain the results of Table 1 that are valid for any $q^2$
and in the limit $m_Q\to\infty$. We notice that in writing the various
contributions in Table 1 we have left the dependence of $p\cdot p^\prime$
on $q^2$, since $p\cdot p^\prime=\frac{1}{2} (m_B^2+m_{K^\star}^2-q^2)$, in the
term arising from the $1^-$ pole and we have assumed that the direct term
has a polar dependence with pole mass given by the $1^+$ pole. These
choices can be justified as follows. The results in Table 1
satisfy, for $q^2\approx q^2_{max}$ the following relations between form
factors of vector and tensor currents:
\bea
A(q^2) &=& i\left\{\frac{q^2-m_B^2-m_{K^\star}^2}{m_B}
\frac{V(q^2)}{m_B+m_{K^\star}}-\frac{m_B+m_{K^\star}}{m_B}A_1(q^2)\right\}
\\
B(q^2) &=& i\frac{2m_B}{m_B+m_{K^\star}}
V(q^2)\\
H(q^2) &=& \frac{2 i}{m_B}\left\{
\frac{V(q^2)}{m_B+m_{K^\star}}+\frac{1}{2q^2}\frac{q^2+m_B^2-m_{K^\star}^2}
{m_B+m_{K^\star}}  A_2(q^2)\right\}
\eea
where $V(q^2)$, $A_j(q^2)$ are the semileptonic form factors in the notations
of ref. \cite{BWS} and they have been obtained in ref. \cite{noi1} by the
same methods employed in the present letter\footnote{In \cite{noi1} we
used $\epsilon^{0123}=+1$ whereas here we use the opposite convention}.
Eqs. (3.1) and (3.2) coincide with the relations found in ref. \cite{isgur};
as for (3.3), the result of \cite{isgur}:
\bea
H(q^2) = \frac{2 i}{m_B}\left\{
\frac{V(q^2)}{m_B+m_{K^\star}}\right.&+&\frac{1}{2q^2}\left(
\frac{q^2+m_B^2-m_{K^\star}^2}
{m_B+m_{K^\star}}  A_2(q^2) \right.\nn \\
&+&\left.\left. 2 m_{K^\star} A_0 (q^2)-(m_{B}+m_{K^\star})
A_1 (q^2) \right) \right\}
\eea
differs from (3.3) for terms that are subleading in the limit $m_Q\to\infty$
and can be neglected. Following \cite{isgur} and \cite{burdman} we assume the
results (3.1)-(3.3) hold also for small values of $q^2$, which justifies the
above mentioned choices in Table 1.
\par
As a final remark we observe that in computing the form factors from Table 1
one has to consider the coupling constants, $\lambda$, $\mu$ and
\be
\alpha_{eff}=\alpha_1(m_P-m_B+m_{K^\star})-\hat\alpha (\frac{\zeta}{2}-
\mu m_{K^\star})
\ee
The constants $\mu$ and $\alpha_{eff}$ have been determined in \cite{noi1} by
an analysis of the $D$ semileptonic decays, with the result
\be
\alpha_{eff}=-0.22\pm 0.02~GeV^{3/2};~~~~~~~~~~~~~~~~
\mu=-0.13\pm 0.05~GeV^{-1}
\ee
A major source of uncertainty in the derivation of $\lambda$ comes from
the value of the leptonic constant $f_D$ or equivalently $\alpha$;
indeed, neglecting $1/m_Q$
corrections (which, incidentally, is consistent with the approximations
made in Table 1), one would obtain, from the value of $|V(0)|$ in
$D\to K^{\star}\ell\nu_{\ell}$ decay:
\be
|\lambda\alpha|=0.16\pm 0.03~GeV^{1/2}
\ee
On the other hand, one could assume a different attitude and take the value
$f_D\approx 200~MeV$ which is suggested by both $QCD$ sum rules
\cite{colangelo2} and the lattice calculations \cite{lattice}.
This amounts to include in the final
results the set of $1/m_Q$ corrections that contribute to $f_D$ and are
known to be large. In this latter case one obtains $|\lambda|=0.60\pm
0.11~GeV^{-1}$. In  \cite{noi1} both cases have been discussed
and the corresponding predictions for $B$ decays were presented. Here
we wish to present an argument in favour of the $scaling$ solution
(3.7).
\par
We can consider the process, related to $B\to K^\star e^+ e^-$,
\be
B\to K^\star\psi
\ee
for which experimental data are available \cite{PDG}:
$BR(B^+\to K^{\star +}\psi)=(1.4\pm 0.7)\times 10^{-3}$ and
$BR(B^0\to K^{\star 0}\psi)=(1.3\pm 0.4)\times 10^{-3}$. Assuming
factorization and using the Wilson coefficients of \cite{BWS2} and
$|V_{cb}|=0.045$ we get, in the case of the $scaling$ solution (3.7),
the result
\be
BR(B\to K^{\star }\psi)=1.5\times 10^{-3}
\ee
which agrees with the data; whereas the solution obtained with the larger
value of $\lambda$ seems disfavoured, since it leads
to $BR(B\to K^{\star }\psi)=2.9\times 10^{-3}$. For this reason we assume
(3.7) as input of our numerical analysis. Using the result (2.16) one would get
$\lambda=0.35\pm 0.06 GeV^{-1}$; in any event, for the calculation of $B\to
K^\star \gamma$, only (3.7) is actually necessary.
\par
In computing the width for the decay $B\to K^\star\gamma$ the combination of
form factors $A+B$ at $q^2=0$ appears. Using Table I we have the result
\be
|A(0)+B(0)|=0.53
\ee
Since the radiative width is given by
\be
\Gamma(B\to K^\star \gamma)=\left(\frac{m_B^2-m^2_{K^\star}}{2m_B}\right)^3
\frac{2|C|^2m_b^2}{\pi}|A(0)+B(0)|^2
\ee
we get,
\be
BR(B\to K^{\star }\gamma)=\left[ 2.5\times \left( |V_{ts}|/0.042 \right)^2
\right]\times 10^{-5}
\ee
where $|V_{ts}|=0.042$ is the central value quoted in \cite{PDG}. In the
previous formula we used $m_t=150~GeV$, $|V_{tb}|\simeq 1$, $m_b=4.7~GeV$ and
$\tau_{B^+}\simeq\tau_{B^0}\simeq\tau_{B_s}\simeq 1.4~ps$,
both for $B^0$ and $B^-$ decays. Our results agree with those of \cite{D1} (in
our notation they find $|A(0)+B(0)| \simeq 0.46$ ), but not with those of
\cite{D2} that are larger than ours by a factor of 2 in the amplitude.
\par
A similar analysis, with obvious changes, applies to the decay
$B_s\to\phi\gamma$. In this case we obtain $|A(0)+B(0)|=0.54$ and
\be
BR(B_s\to \phi\gamma)=\left[2.7 \times \left( |V_{ts}|/0.042 \right)^2
\right]\times 10^{-5}
\ee
Taking only into account the experimental uncertainty of $V_{ts}$ gives
for the branching ratio
$B\to K^{\star }\gamma$ a range from $1.3 \times 10^{-5}$ to $4.1 \times
10^{-5}$, and for $B_s\to \phi\gamma$ from $1.4 \times 10^{-5}$ to $4.5 \times
10^{-5}$.
\resection{$B\to K e^+ e^-$ and $B\to K^\star e^+ e^-$}
These decays occur dominantly via a quark process $b\to s\gamma^\star\to
s e^+ e^-$ ($\gamma^\star=$ virtual photon). In the effective lagrangian
for $b\to s e^+ e^-$ we have to include also the so-called long-distance
contributions arising from $\psi-\gamma$ or $\psi^\prime-\gamma$
conversion, i.e. from the quark subprocess $b\to s\psi\to s e^+ e^-$.
The effective lagrangian has been derived in \cite{inami}, \cite{nn}
and we shall not report it here for the sake of simplicity.
\par
Let us first discuss $B\to K e^+ e^-$. The short distance hadronic matrix
element relative to this decay has been given in (2.6). We compute it by
using the lagrangian and currents of Section 2. The corresponding diagrams
are similar to those of $B \to K^\star \gamma$, with $K^\star$ changed into
$K$ and the pole given by the $1^-$ $\bar s b$ meson; however it turns out that
there is no direct coupling and the only surviving term, in the limit
$m_Q\to\infty$, and for $q^2\approx q^2_{max}$, is the polar contribution.
\par
Assuming, as in previous case, a $q^2$ dependence given by a simple pole
(with $m_P=m_{B^\star_s}$), we get the result
\be
S(q^2)=\frac{S(0)}{1-q^2/m^2_{B_s^\star}}
\ee
with
\be
S(0)=\frac{\alpha g }{f_\pi m^2_{B^\star_s}\sqrt{m_B}}(m_{B^\star_s}+
m_B-m_K)
\ee
We can express this result in terms of the form factors which appear
in the matrix element of the $V-A$ current ($F_1(0)=F_0(0)$)
\be
\langle K(p^\prime)|\bar s\gamma^\mu (1-\gamma_5) b|{\bar B}(p)\rangle
=[p^\mu+p^{\prime\mu}+\frac{m_K^2-m_B^2}{q^2} q^\mu]F_1(q^2)
-\frac{m_K^2-m_B^2}{q^2} q^\mu F_0(q^2)
\ee
and have been computed in \cite{noi1}. The following relation
between the form factors holds:
\be
S(q^2)=-\frac{2F_1(q^2)}{m_B}
\ee
\par
Two remarks are in order. First (4.4) coincides with the analogous
relation found in \cite{isgur} by Isgur and Wise
\be
S(q^2)=\frac{1}{m_B}\left[-F_1(q^2)+\frac{m^2_K-m^2_B}{q^2}\left(F_1(q^2)-
F_0(q^2)\right)\right]
\ee
only at $q^2\approx q^2_{max}$ and $m_B\to\infty$. As in the case of
equation (3.3)
we find that some form factors (in this case $F_0$) are subleading
when $m_Q\to\infty$, which is expected because the $0^+$ state,
contributing to $F_0$, cannot couple to the antisymmetric tensor
current $\bar s\sigma_{\mu\nu}(1+\gamma_5) b$. The second
remark is related to the value of the coupling constant $g$ in (4.2).
One can obtain it by using data on $D\to K\ell\nu_{\ell}$ or
$D\to \pi\ell\nu_{\ell}$, together with flavour symmetry and one meets
the same problem found in discussing the constant $\lambda$ in
the previous Section. Using the scaling hypothesis we get from
semileptonic data \cite{stone}
\be
|g\alpha|=0.17\pm 0.06~GeV^{3/2}
\ee
which therefore gives
\be
|S(0)|=0.18 ~GeV^{-1},~~~~~~~~~~~~~|F_1^{B\to K}(0)|=0.5
\ee
Incidentally we note that, using for $\alpha$ the information from QCD sum
rules (2.16), one obtains for $g$ in the scaling hypothesis $|g|=0.37\pm 0.13$
but again only (4.6) is necessary for $B\to K e^+e^-$.
\par
Alternatively one can substitute $\alpha/\sqrt{m_D}$ with $f_D\approx
200~MeV$ in $|F_1^{D\to \pi}(0)|$; this would lead to
$|F_1^{B\to K}(0)|=0.85$. As in  the previous case one can get a suggestion
on how to solve this ambiguity from the non leptonic decay
\be
B\to K\psi
\ee
\par
The experimental data are \cite{PDG}:
$BR(B^+\to K^{+}\psi)=(7.7\pm 2.0)\times 10^{-4}$ and
$BR(B^0\to K^{ 0}\psi)=(6.5\pm 3.1)\times 10^{-4}$; they are compatible
with the $scaling$ hypothesis, i.e. with the choice represented by
the eqs. (4.6)-(4.7). As a matter of fact (4.7) gives, together with
the factorization approximation and $|V_{cb}|=0.045$,
\be
BR(B^+\to K^{+}\psi)=BR(B^0\to K^{ 0}\psi)=1.1\times 10^{-3}
\ee
whereas with the non scaling assumption ($|F_1^{B\to K}(0)|=0.85$)
one would get for the branching ratio the value $3.2\times 10^{-3}$,
which is excluded by the data. Even if this argument is based on additional
hypotheses, we take it as a strong indication in favour of the scaling
behaviour
of the form factors.
\par
Using (2.6) and (4.1)-(4.7), together with the effective lagrangian for the
process $b\to s e^+ e^-$, including long distance contributions \cite{inami},
\cite{nn}, we can get
the distribution $d\Gamma(B\to K e^+ e^-)/dQ^2$ in the invariant mass
squared of the lepton pair $Q^2$. We can repeat the same analysis for
$d\Gamma(B\to K^\star e^+ e^-)/dQ^2$, using the form factors
reported in Section 3 together with the weak $V-A$ current given in
\cite{noi1}. These results are reported in Figs. 1-2. We confirm the results
obtained by previous authors \cite{riaz}, \cite{deshpande}, showing that
the widths for the processes $B\to K e^+ e^-$ and $B\to K^\star e^+ e^-$
are largely dominated by the long distance contributions
$B\to K^{(\star)} \psi , \psi^\prime \to K^{(\star)} e^+ e^-$.
Nevertheless an accurate measurement of the lepton pair spectrum below
$c\bar c$ resonances would display the effects of the short distance dynamics
arising from the hamiltonian (2.1). This measurement would therefore
complement the analysis of the $B\to K^\star\gamma$ decay process
providing further information on the fundamental parameters appearing
in (2.1).
\par
We conclude this letter by giving the branching ratios for the decay
$B\to K\psi(2S)$ and $B\to K^\star\psi(2S)$ that can be obtained as
byproduct of our analysis
\be
BR(B^+\to K^{+}\psi(2S))=BR(B^0\to K^{ 0}\psi(2S))=3.9\times 10^{-4}
\ee
\be
BR(B^+\to K^{\star +}\psi(2S))=BR(B^0\to K^{\star 0}\psi(2S))=8.0\times 10^{-4}
\ee
These results are within the experimental upper bounds quoted in \cite{PDG},
which are $1.5 \times 10^{-3}$ for the branching ratio (4.10) and $3.5 \times
10^{-3}$ for (4.11).
\resection{Conclusions}
Radiative decays of the $B$ mesons possess an important potential for exploring
certain elements of the Standard Model and also for discovering possible new
physics. We have employed effective chiral theory including mesons with one
heavy quark to calculate the decays $B \to K^{\star}\gamma$ and $B_s \to \phi
\gamma$, and the pair production processes $B \to K e^+ e^-$ and $B \to
K^{\star} e^+ e^-$. The inherent symmetries have allowed us to calculate these
processes in terms of some constants previously determined from the study of
$B$
and $D$ semileptonic decays within the same model. Extrapolation from the
region of zero recoil momentum has required particular attention. Our final
result for photon decays is
\bea
10^5 \times BR(B \to K^{\star} \gamma) &=& 2.5 \times \left( \frac{|V_{ts}|}
{0.042} \right)^2 \\
10^5 \times BR(B_s \to \phi \gamma) &=& 2.7 \times \left( \frac{|V_{ts}|}
{0.042} \right)^2
\eea
Present errors in $V_{ts}$ give for $BR(B \to K^{\star}\gamma)$
values from $1.3 \times 10^{-5}$ to $4.1 \times 10^{-5}$,
 and for $BR(B_s \to \phi \gamma)$ from
$1.4 \times 10^{-5}$ to $4.5 \times 10^{-5}$. Additional uncertainties are of
course implicit in the model chosen, and we have discussed them in detail.
Concerning $B \to K e^+ e^-$ and $B \to K^{\star} e^+ e^-$, long distance
contributions have to be included, which we take as dominantly given by
$B \to K^{\star} \psi$ and $B \to K^{\star} \psi'$, and subsequent $\psi$,
$\psi'$ conversion into $\gamma$. Comparison with the predicted lepton pair
mass distributions in their kinematical ranges would allow for verification of
both short distance  and long distance terms.
\par
\vspace*{1cm}
\noindent
{\bf  Acknowledgement}:
Useful discussions with Dr. P.Colangelo, Dr. F.Feruglio and Prof. N.Paver are
gratefully acknowledged. We also thank Prof. S.Stone for kindly communicating
us the CLEO result.
\par
\vspace*{1cm}
\noindent
{\bf NOTE ADDED}: After having completed this work results from CLEO
\cite{Stone} where made public, giving  $BR(B \to K^{\star}\gamma)=(4.5 \pm
1.5 \pm 0.9) \times 10^{-5}$, which agrees, within the error, with our
prediction (3.12).

\newpage
\noindent
{\bf Table I}. Terms contributing to the various form factors of the transition
$B\to K^\star\gamma$. $m_P$ is the pole mass ($m_P=5.71~GeV$ for the direct
and $1^+$ term; and $5.32~GeV$ for the $1^-$ contribution). $p\cdot p^\prime=
(m_B^2+m_{K^\star}^2-q^2)/2$.
\begin{table}
\begin{center}
\begin{tabular}{l c c c }
\hline \hline
${\rm Form~Factor}$ & ${\rm Direct}$ & $1^-$ & $1^+$  \\ \hline\\
$A(q^2)$ &$\displaystyle{
\frac{i\sqrt{2}g_V\alpha_1}{\sqrt{m_B}}\frac{q^2_{max}-m_P^2}
{q^2-m_P^2}}$ &$\displaystyle{
\frac{i2\sqrt{2}\alpha\lambda g_V(p\cdot p^\prime)}
{(m_P^2-q^2)\sqrt{m_B}}}$ &$\displaystyle{
\frac{-i\sqrt{2m_B}\hat\alpha g_V(\zeta-2\mu
m_{K^\star})}{m_P^2-q^2}}$ \\\\ \hline\\
$B(q^2)$ &0 &$\displaystyle{
\frac{-i2\sqrt{2}\alpha\lambda g_V m_B^{3/2}}{m_P^2-q^2}}$
 & 0 \\\\ \hline\\
$H(q^2)$ & 0&$\displaystyle{
\frac{-i2\sqrt{2}\alpha\lambda g_V}{(m_P^2-q^2)\sqrt{m_B}}}$ &
$\displaystyle{
\frac{-i2\sqrt{2m_B}\hat\alpha g_V \mu}{(m_P^2-q^2)m_B}}$ \\\\ \hline
\hline
\end{tabular}
\end{center}
\end{table}
\newpage
\begin{center}
  \begin{Large}
  \begin{bf}
  Figure Captions
  \end{bf}
  \end{Large}
  \vspace{5mm}
\end{center}
\begin{description}
\item [Fig. 1] Differential distribution in the invariant mass squared of the
lepton pair $q^2$ for the process $B \to Ke^+e^-$.
\item [Fig. 2] Differential distribution in the invariant mass squared of the
lepton pair $q^2$ for the process $B \to K^*e^+e^-$.
\end{description}

\newpage
\begin{thebibliography}{99}
\bibitem{bertolini}
S.Bertolini, F.Borzumati and A.Masiero, Phys. Rev. Lett. {\bf 59} (1987)
180; N.G.Deshpande, P.Lo, J.Trampetic, G.Eilam and P.Singer, Phys. Rev.
Lett. {\bf 59} (1987) 183; B.Grinstein, R.Springer and M.B.Wise,
Nucl. Phys. {\bf B339} (1990) 269; R.Grigjanis, P.J.O'Donnell,
M.Sutherland and H.Navelet, Phys. Lett. {\bf B237} (1990) 355;
G.Cella, G.Curci, G.Ricciardi and A.Vicer\'e, Phys. Lett. {\bf B248}
(1990) 181; M.Misiak, Phys. Lett. {\bf B269} (1991) 161.
\bibitem{vasanti}
N.Vasanti, Phys. Rev. {\bf D13} (1976) 1889; M.A.Shifman, A.I.Vainshtein
and V.I.Zakharov, Phys. Rev. {\bf D18} (1978) 2583; Y.I.Kogan and
M.A.Shifman, Sov. J. Nucl. Phys. {\bf 38} (1983) 628.
\bibitem{inami}
T.Inami and C.S.Lim, Progress Theor. Phys. {\bf 65} (1981) 297.
\bibitem{bertolini2}
S.Bertolini, F.Borzumati and A.Masiero, Phys. Lett. {\bf B192} (1987)
437; R.Barbieri and G.F.Giudice, preprint CERN-TH.6830/93.
\bibitem{isgur}
N.Isgur and M.B.Wise, Phys. Rev. {\bf D42} (1990) 2388.
\bibitem{gen}
N.Isgur and M.B.Wise, Phys. Lett. {\bf B232} (1989) 113; ibidem {\bf B237}
(1990) 527; M.B.Voloshin and M.A.Shifman, Sov.J.Nucl.Phys. {\bf45} (1987)
292; ibidem {\bf 47} (1988) 511; H.D.Politzer and M.B. Wise, Phys. Lett.
{\bf 206B} (1988) 681; ibidem {\bf 208B} (1988) 504; E.Eichten and B.Hill,
Phys.
Lett. {\bf 234B} (1990) 511; H.Georgi, Phys.Lett. {\bf 240B} (1990) 447;
B.Grinstein, Nucl. Phys. {\bf B339} (1990) 253; A.F.Falk, H.Georgi, B.Grinstein
and M.B.Wise, Nucl. Phys. {\bf B343} (1990) 1.
\bibitem{burdman}
G.Burdman and J.F.Donoghue, Phys. Lett. {\bf B270} (1991) 55.
\bibitem{donnell}
P.J.O'Donnell and H.K.K.Tung, preprint UTPT-93-02.
\bibitem{wise}
M.B.Wise, Phys. Rev. {\bf D45} (1992) R2188.
\bibitem{noi1}
R.Casalbuoni, A.Deandrea, N.Di Bartolomeo, F.Feruglio, R.Gatto and
G.Nardulli, Phys. Lett. {\bf B299} (1993) 139; ibidem {\bf B292} (1992) 371.
\bibitem{riaz}
C.A.Dominguez, N.Paver and Riazuddin, Z. Phys. {\bf C48} (1990) 55.
\bibitem{deshpande}
N.G.Deshpande, J.Trampetic and K.Panose, Phys. Lett. {\bf B214} (1988) 467.
\bibitem{colangelo}
P.Colangelo, G.Nardulli, N.Paver and Riazuddin, Z. Phys. {\bf C45} (1990)
575.
\bibitem{colangelo2}
C.A.Dominguez and N.Paver, Phys. Lett. {\bf B197} (1987) 423, E {\bf B199}
(1987) 596;
P.Colangelo, G.Nardulli, A.A.Ovchinnikov and N.Paver, Phys. Lett. {\bf B269}
(1991) 201; L.J.Reinders, Phys. Rev. {\bf D38} (1988) 947.
\bibitem{broadhurst}
D.J.Broadhurst and A.G.Grozin, Phys. Lett. {\bf B267} (1991) 105 and {\bf B274}
(1992) 421; M.Neubert, Phys. Rev. {\bf D45} (1992) 2451;
P.Colangelo, G.Nardulli and N.Paver, preprint BARI-TH 9/132, UTS-UFT-93-3
(1993).
\bibitem{colangelo3}
P.Colangelo, G.Nardulli and N.Paver, Phys. Lett {\bf B293} (1992) 207.
\bibitem{stone}
S.Stone, Syracuse University report HEPSY-1-92, in Heavy Flavours,
A.J.Buras and H.Lindner eds. (Singapore 1992).
\bibitem{BWS}
M.Wirbel, B.Stech and M.Bauer, Z. Phys. {\bf C29} (1985) 637.
\bibitem{lattice}
A.Abada et al., Nucl. Phys. {\bf B376} (1992) 172.
\bibitem{PDG}
Particle Data Group, Review of Particle Properties, Phys. Rev. {\bf D45}
(1992) S1.
\bibitem{BWS2}
M.Bauer, B.Stech and M.Wirbel, Z. Phys. {\bf C34} (1987) 103.
\bibitem{D1}
N.G.Deshpande, P.Lo, J.Trampetic, Z.Phys. {\bf C40} (1988) 369.
\bibitem{D2}
C.A.Dominguez, N.Paver and Riazuddin, Phys. Lett. {\bf B214} (1988) 459.
\bibitem{nn}
N.G.Deshpande, Bombay HEP Workshop 1989 pgs.538-558, talk given at the
workshop on High Energy Phenomenology TIFR, Bombay (1989).
\bibitem{Stone}
S. Stone, private communication.
\end{thebibliography}
\end{document}